# Parametric shape optimization for combined additive-subtractive manufacturing

Authors:

1. Christian Altenhofen (a,b)
2. Marco Attene (c)
3. Oliver Barrowclough (d)
4. Michele Chiumenti (e)
5. Marco Livesu (c)
6. Federico Marini (f)
7. Massimiliano Martinelli (f)
8. Vibeke Skytt (d)
9. Lorenzo Tamellini (corresponding author, tamellini@imati.cnr.it) (f)

Affiliations:

a. Fraunhofer Institute for Computer Graphics Research IGD, Interactive Engineering Technologies, Darmstadt, Germany
b. Technische Universität Darmstadt, Interactive Graphics Systems Group, Darmstadt, Germany
c. Consiglio Nazionale Delle Ricerche, Istituto di Matematica Applicata e Tecnologie Informatiche (CNR-IMATI), Genova, Italy
d. SINTEF Digital, Mathematics and Cybernetics, Oslo, Norway
e. International Center for Numerical Methods in Engineering (CIMNE) Universidad Politécnica de Cataluña, Barcelona, Spain.
f. Consiglio Nazionale Delle Ricerche, Istituto di Matematica Applicata e Tecnologie Informatiche (CNR-IMATI), Pavia, Italy

## Abstract

In the industrial practice, additive manufacturing processes are often followed by post-processing operations such as subtractive machining, milling, etc. to achieve the desired surface quality and dimensional accuracy. Hence, a given part must be 3D printed with extra material to enable such finishing phase. This combined additive/subtractive technique can be optimized to reduce manufacturing costs by saving printing time and reducing material and energy usage.

In this work, a numerical methodology based on parametric shape optimization is proposed for optimizing the thickness of the extra material, allowing for minimal machining operations while ensuring the finishing requirements. Moreover, the proposed approach is complemented by a novel algorithm for generating inner structures leading to reduced distortion and improved weight reduction. The computational effort induced by classical constrained optimization methods is alleviated by replacing both the objective and constraint functions by their sparse-grid surrogates. Numerical results showcase the effectiveness of the proposed approach.

## 1.1 Keywords

Additive manufacturing, shape optimization, parametric optimization, sparse grids, surrogate model

# 1 Introduction

The work described in this paper has been developed as part of the Horizon 2020 research and innovation project CAxMan (*Computer Aided technologies for Additive Manufacturing*, see https://www.caxman.eu). The objectives of this project were to establish cloud-based toolboxes, workflows, as well as one-stop shop for CAx technologies supporting the design, simulation and process planning for Additive Manufacturing. The project originates from the observation that, due to the constant growth of the additive manufacturing (AM) market, there is an increasing demand for a software ecosystem that enables Computer Aided Technologies (CAx) support of AM processes and machines. However, to move from prototypes and demonstrators to actual industrial use-cases, the final quality of the AM products, such as strength, surface quality, metallurgy and dimensional control, must be ensured. Thus, the project established novel cloud-based workflows and services for discrete manufacturing (combinations of additive and subtractive) by addressing analysis/simulation-based design and analysis/simulation of process-planning, showcasing the CAx technologies for two use-cases from design to production. One of these workflows was the shape optimization procedure detailed in this paper.

Indeed, one of the major reasons for the success of additive manufacturing (AM) is because it enables the fabrication of very complex geometries out of reach with standard manufacturing processes, such as milling, turning or casting. However, the quality of AM parts is often not sufficient for industrial applications due to thermal distortions induced by the process and surface roughness. To enhance AM quality and to achieve the required dimensional accuracy, a possible solution consists of adding some extra material (geometric *offset*) in order to enable the part to be finished by subtractive manufacturing. This solution introduces two limitations: on the one hand, if too little extra material is added the surface roughness cannot be removed adequately; on the other hand, adding too much material unnecessarily increases printing time and manufacturing costs. Note that the original dimensions of the *nominal* geometry are compromised by the distortions accumulated during the AM process due to the thermal stresses. Hence, the input for the 3D-printer is not the nominal geometry but the so-called *stock* geometry instead. In addition to the geometric offset, we also consider the possibility to add inner cavities to the stock part, for saving weight and production time/costs. The generation of inner structures also affects the structural behavior of the entire part, not only regarding the in-service requirements, but also during the manufacturing phases. In fact, reducing the amount of material to be sintered as well as reducing the surface area per layer modifies the thermal behavior in terms of heat accumulation and heat extraction.

Therefore, the main objective of this workflow was the definition of a numerical procedure for the optimization of the stock part design to be provided as input to a metal 3D-printer, keeping into account the thermal deformation induced by the AM process itself and the following post-production by subtractive machining. The problem can be defined as a multi-dimensional constrained optimization analysis, that can be solved using classical optimization methods, see e.g. [1]. While many works are available in literature on generic shape optimization [2] [3] [4], less works are available on printer-aware optimization [5] [6] [7] [8] [9]; this is therefore one of the major contributions of this work.

The optimization method requires the numerical simulation of the AM process for a number of different stock part designs. This kind of software platform needs to compute the residual stresses and distortions induced by the thermal deformations due to the AM process including the final cooling phase and, eventually, the post-treatment processes [10] [11]. These analyses are very CPU-time demanding requiring a software platform based on massive parallel computing via sub-domain decomposition methods, octree-based meshing tools and embedded technologies [12]. Hence, to alleviate this inconvenience, a surrogate model for both the objective and constraint functions is adopted in the optimization problem; specifically, a sparse grid method has been adopted in this work [13] [14] [15].

Another novel aspect of this work is that we strived for an as-generic-as-possible and black-box implementation, that takes as input a standard STEP file and seamlessly runs through the different pieces of software (provided by different institutions involved in the project) that are needed for the generation of stock parts, the AM process simulation and the outer optimization loop. In particular, we implemented the methodology in a cloud-based environment in which the single pieces of software are run as black boxes in a pipeline (i.e., the output of a software is the input of the following software).

The content of this paper is organized as follows. Section 2 discusses the overall methodology by focusing on a specific test-case for the sake of clarity (the generalization to more complex use-cases is straightforward). This includes discussion of the specifications of the test-case (Section 2.1), the optimization methodology (Section 2.2), the corresponding numerical techniques used for its resolution (Section 2.3), the definition of the surrogate model (Section 2.4) and the different steps for the evaluation of the objective and constraint functions according to the stock design parametrization (Section 2.5). We then give some implementation details in Section 2.6. Section 3 presents some numerical examples that demonstrate the performance of the optimization algorithm. Finally, Section 4 summarizes the work, presenting some new perspectives and possible directions for future work.

## 2 Methodology

### 2.1 The industrial use case

The mock-up test case considered in this work is based on the industrial model of a gear, selected within the CAxMan project and provided by the high-tech engineering company STAM, based in Genoa, Italy (http://www.stamtech.com). Figure 1-left shows the nominal geometry of the gear. The final goal is the design optimization ready for the printing process.

For the sake of simplification, a single gear tooth as extracted by the original geometry will be used to show the optimization strategy proposed in this work (Figure 1 – right). Note that the application to the entire gear is straight-forward but the computational time required by the numerical simulation of the AM process was very high and useless to illustrate the chosen methodology.

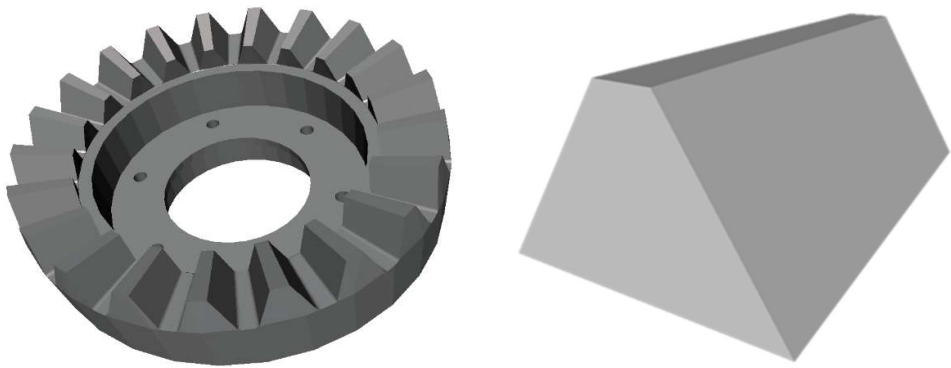

*Figure 1: Gear model provided by STAM (left) and an approximation of a single tooth representing the "nominal geometry" selected (right)*

### 2.2 The constrained optimization formulation

The objective is the optimization of the stock part geometry to minimize the subtractive post-processing work after the AM stage. The stock part of our mock-up test is parametrized by 3 parameters: the first one is the offset thickness while the other two parameters are used to control the generation of inner structures as described in Section 2.5.1. These parameters are collected in the vector $\mathbf{p} \in \Gamma \subset \mathbb{R}^3$. The goal is finding the design variables that minimize the volume of the extra material needed, while at the same time allowing for the machining operations after the distortions induced by the AM process and the post-processing operations.

Mathematically, this writes as:

$$min_{\mathbf{p}\in\Gamma}\Delta Volume(\mathbf{p})$$
$$st.\Delta Thickness(\mathbf{p}) > 0.04mm$$
$$\Gamma = [p_{1,min}, p_{1,max}] \times [p_{2,min}, p_{2,max}] \times [p_{3,min}, p_{3,max}]$$

where:
- $\Delta Volume(\mathbf{p})$ is the total volume of the extra material of the stock part, computed as the difference between the volumes of the nominal and the stock geometries;
- $\Delta Thickness(\mathbf{p})$ is the thickness of the material that must be removed from each surface after the 3D printing. Because of the machining tolerance, this thickness must be larger than 0.04mm.
- $\Gamma = [p_{1,min}, p_{1,max}] \times [p_{2,min}, p_{2,max}] \times [p_{3,min}, p_{3,max}]$ is the Cartesian product defining the range of variability of the 3 parameters defining the stock-part: the so called *constraints box*.

The proposed problem can easily be generalized by assuming a different offset thickness and machining tolerance for each surface of the geometry: the resolution method proposed here would still be applicable.

### 2.3   Numerical methods for constrained optimization

The constrained optimization problem thus reads as:

$$\mathbf{p}^* = \arg\min_{\mathbf{p}\in\Gamma} f(\mathbf{p})$$
$$st: g(\mathbf{p}) \leq 0$$

Where $f(\mathbf{p}) = \Delta Volume(\mathbf{p})$ and $g(\mathbf{p}) = \Delta Thickness(\mathbf{p})$. In this work, the penalization method [1] is used to transform the initial constrained problem into the following modified unconstrained one:

$$\mathbf{p}^* = \arg\min_{\mathbf{p}\in\mathbb{R}^N} f(\mathbf{p}) + \tilde{g}(\mathbf{p}), \quad \mathbf{p} \in \mathbb{R}^N$$

where $\tilde{g}(\mathbf{p}) \gg 0$ if $g(\mathbf{p}) > 0$ to avoid unfeasible solutions. Several methods to build the function $\tilde{g}(\mathbf{p})$ and to solve the unconstrained optimization problem are available in literature; we will discuss them later in this section. Before that however we need to make an important remark: evaluating $f(\mathbf{p}), g(\mathbf{p}), \tilde{g}(\mathbf{p})$ require calling an expensive FEM solver to estimate the distortion of the stock geometry. Since the optimization method requires multiple evaluations, in this work these functions are replaced by surrogate models (i.e., approximating functions), which are built beforehand by employing a handful of full evaluations of $f(\mathbf{p}), g(\mathbf{p}), \tilde{g}(\mathbf{p})$ only, and are very cheap to evaluate. We denote these surrogate models by capital letters, $F(\mathbf{p}), G(\mathbf{p}), \tilde{G}(\mathbf{p})$; the optimization problem that we eventually solve is therefore

$$\mathbf{p}^* = \arg\min_{\mathbf{p}\in\mathbb{R}^N} F(\mathbf{p}) + \tilde{G}(\mathbf{p}), \quad \mathbf{p} \in \mathbb{R}^N \qquad (1.1)$$

Most of the CPU-time will actually be needed to perform the preliminary numerical simulation of the AM process to construct these surrogate models, while solving the optimization problem (1.1) is fast. Therefore in this work we do not restrict to a single choice of penalization function $\tilde{G}(\mathbf{p})$ and of a single method for unconstrained optimization. Instead, we employ three penalization methods and two unconstrained optimization methods (so, six combinations in total), and choose finally the best solution proposed by all these methods. The penalization methods used are the squared penalty method, the augmented Lagrangian and the log-barrier method. The two unconstrained optimization methods are the gradient-descent and the Nelder-Mead methods: the former is faster but might fail if the optimal point is close to the boundary of the constraint box, while the latter is slower but more robust. Finally, different initial conditions chosen according to random Latin Hypercube Sampling method have been considered, to reduce the risk of stagnation in local minima [16].

## 2.4 Surrogate model construction

A vast body of literature on surrogate models construction is available: radial basis functions, reduced basis, proper orthogonal decomposition, neural networks, sparse grids among can be used for the scope, see e.g. [17] [18] [19] [20] [21]. In this work, the sparse grid method [13] [14] [15] is chosen. This method is suitable to approximate multivariate function depending on a moderate number of inputs (say, up to a dozen), and has been successfully employed as surrogate model in optimization problems, see e.g. [22]. It is easy to use, since it only requires evaluating the functions at hand at some prescribed points, and works well if the function to be approximated is smooth. An example of set of evaluation points over a two-dimensional constraint box $\Gamma$ is shown in Figure 2; each black dot represents a point $\mathbf{p}$ for which $f(\mathbf{p})$ and $g(\mathbf{p})$ must be evaluated to construct the surrogate models $F(\mathbf{p}), G(\mathbf{p}), \tilde{G}(\mathbf{p})$.

In the most basic form of sparse grids (the one that we use), the number of evaluation points cannot be prescribed by the user. Instead, the user prescribes an integer number *w*, which is the "refinement level".

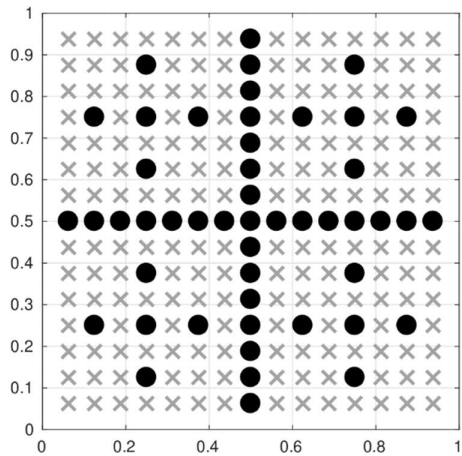

*Figure 2: Sparse grid over the domain [0,1]^2. The sparse grid is a subset of a full Cartesian grid (points in grey crosses).*

The higher w, the more points in the sparse grids and the more accurate the surrogate model. Table 1 reports the number of points in a sparse grids for several values of N (number of optimization parameters) and *w*.

| N | w=1 | w=2 | w=3 | w=4 | w=5 |
|---|---|---|---|---|---|
| 1 | 1 | 3 | 7 | 15 | 31 |
| 2 | 1 | 5 | 17 | 49 | 129 |
| 3 | 1 | 7 | 31 | 111 | 351 |

*Table 1: number of sample points for a given number of optimization parameters and different levels w. The grid at N=2, w=4 is the one shown in Figure 2.*

In summary, the optimization procedure proceeds are follows:

For *w*=1, 2, …

1. fix the sparse grid level *w* and determine the sparse grid points;
2. for each point, evaluate the constraint and objective function for a fixed stock design;
3. build the sparse grid surrogate model for constraint and objective functions;
4. solve the constrained optimization problem obtained by replacing the constraint and objective problem with their sparse grid surrogates, i.e., equation (1.1);
5. if the results are not satisfactory (e.g., if the optimal point has changed more than a given tolerance with respect to the one obtained ad level *w*-1), increase the sparse grid level and repeat.

Observe that Point 2 can be performed in parallel, as the evaluations are independent from each other. Moreover, it is worth pointing out that increasing the sparse grid level, generates a new grid which typically includes the previous one, so that all computations already performed can be recycled. In the following, the evaluation of the constraint and objective functions for a fixed stock design is discussed.

## 2.5 Evaluation of objective and constraint function

Evaluating the objective and constraint function for a given stock design is a complex, multi-stage operation which requires the use of different software. Upon fixing the values of the design parameters **p** for the stock design, the following steps are needed:

1. Generation of inner cavities and structures;
2. Construction of the offset geometry starting from the definition of the nominal geometry and the inner cavities;
3. Mesh generation;
4. Simulation of the AM process;
5. Computation of the distance between deformed stock design and nominal geometry, i.e., the constraint function $\Delta Thickness(\mathbf{p})$;
6. Computation of the volume of the difference between the stock and the nominal geometry, i.e., the objective function $\Delta Volume(\mathbf{p})$.

### 2.5.1 Generation of inner cavities

In order to reduce the distortion during the printing process, internal cavities inside the nominal geometry can be generated. In fact, reducing the volume of the part mitigates its thermal deformation. Moreover, introducing such cavities leads to a weight reduction saving material and printing time. A major assumption that we make is that the generation of inner cavities is compatible with structural/mechanical performance of the component to be printed. If this is not the case, it is always possible to increase the complexity of the optimization problem adding any further mechanical constraints.

Internal cavities are created by fitting a repetitive pattern based on a regular hexahedral grid inside the bounding box of the nominal geometry. Different patterns hold different thermo/mechanical behaviors in terms of heat diffusion, structural stiffness and material savings. The intersection (Boolean operation) between the volume of the nominal geometry and the structure chosen for the inner cavities leads to the final geometry of the component.

The geometries of the inner structures are based on Catmull-Clark (CC) subdivision surfaces [23]. This technique is commonly used in computer graphics and animation, and more and more used in the engineering sector. CC subdivision allows for defining smooth surfaces ($C^2$-continuous in regular areas, $C^1$-continuous around extraordinary points) based on a discrete control mesh. The so-called *limit surface* can be created iteratively, by repeatedly applying (ideally, infinitely many times) the iterative subdivision steps, or can be directly evaluated numerically [24]. Because the regular CC limit patches correspond to bi-cubic B-spline patches, CC subdivision surfaces, and therefore also the geometry of our internal structures, can be converted into a CAD-compatible representation as well. However, irregular CC patches must be approximated, introducing a slight approximation error during the conversion process. To create the internal structures, all the faces that separate an *empty* grid cell from a *filled* one are combined into one or multiple closed CC subdivision control mesh.

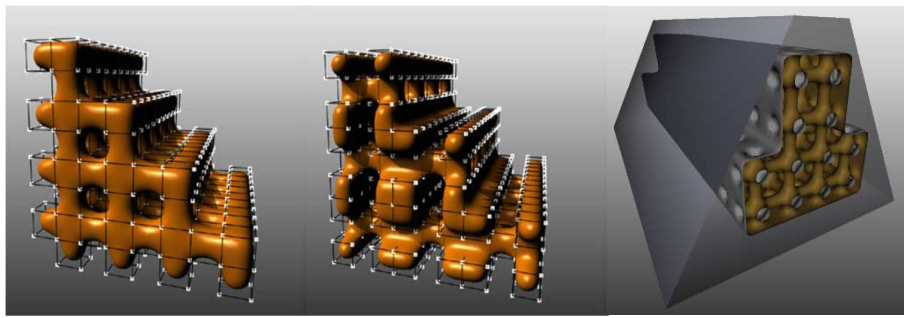

*Figure 3: Three different possible patterns for internal cavities; the rightmost picture shows the internal structures inserted in the nominal geometry*

Figure 3 shows three possible patterns repeated in a grid-based manner. By alternating different patterns, it is possible to treat different regions of the part according to local structural requirements. Note that 90º angles between grid cells result in curved geometry due to the smoothing property of CC surfaces. This is beneficial in terms of manufacturability because modern AM machines (especially SLS and SLM machines) are able to manufacture overhangs of up to 45 degrees, unless the geometry has a curved, arch-like shape. Therefore, the overall quality of the inner cavities is quite good in terms of surface quality and dimensional control. In this work, the so-called "grid" pattern is adopted as shown in the rightmost picture in Figure 3. The generation depends on two optimization parameters:

1. The grid resolution (size of the cavities);
2. The minimum wall thickness between the cavities and the outer surface of the object.

Both parameters are used to define the parameter space in which the optimization is performed. Table 2 shows some examples of inner structures created varying wall thickness from 0.0 mm to 0.5 mm. It can be seen that the number of repetitions, as well as the amount of material saved, reduce as the wall thickness is increased.

| Wall thickness | 0.0 mm | 0.1 mm | 0.2 mm | 0.3 mm | 0.4 mm | 0.5 mm |
|---|---|---|---|---|---|---|
| Resulting cavities | 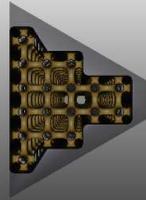 | 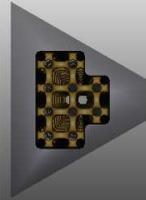 | 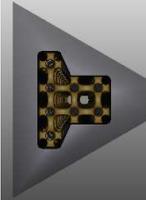 | 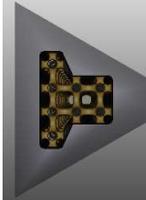 | 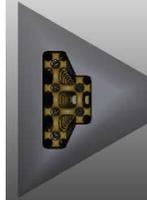 | 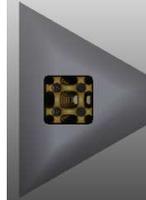 |
| Remaining material | 50.89 % | 74.82 % | 80.02 % | 81.57 % | 86.27 % | 90.88 % |

*Table 2: effect of the parameter "wall thickness" for generation of inner structures.*

### 2.5.2   Geometry offset and mesh generation

The offset of a three-dimensional object, *S*, can be conceptualized as follows: let *d(x)* be the distance function from the closest point in *S* (thus, a function $R^3 \to R$ such that *d(x)* = 0 if **x** $\in$ *S*). Then, *offset$_r$(S)* = {**x** $\in$ $R^3$ s.t. *d(x)* < *r*} is the offset of *S* at radius *r*. Hence, a point **x** belongs to the *Offset$_r$(S)* if and only if **x** is either in *S* or in an *r*-radius sphere centered on its surface.

In this work, starting with an initial tessellated geometry, whose vertices belong to the nominal geometry, the objective is to obtain an "inflated" geometry satisfying the required allowance. Unfortunately, a plain offsetting cannot be used because the class of polyhedra is not closed under the offsetting operation. Therefore, the exact offset of a polyhedron may have parts of its surface that are not piecewise-linear (i.e. some surface parts may be locally cylindrical or spherical). Exact offsetting might therefore only be possible for a very small set of simple geometries.

By observing that an offset is a particular case of a so-called *Minkowski sum*, another solid *M* can be used to obtain an arbitrary Minkowski sum instead of rolling a sphere over the surface of *S*; the result is denoted by *S*⊕*M*. Thus, if *M* is a sphere with radius *r*, the Minkowski sum *S*⊕*M* is equal to *Offset$_r$(S)*. Figure 4 shows an example of the Minkowski sum of a spoon and a star model.

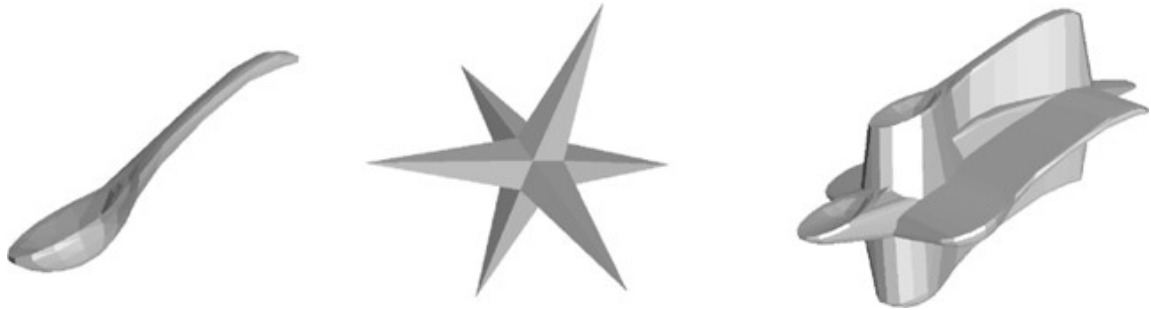

*Figure 4: Minkowski sum of a spoon and a star (image courtesy of Peter Hachenberger and the CGAL project).*

Note that the Minkowski sum of two polyhedra is a polyhedron. Thus, we exploited this result to implement an approximated offsetting whose result is a polyhedron with guaranteed distance bounds from *S*. Our implementation uses the CGAL library [25] and the concept of NEF polyhedron [26]. In the experiments reported in this paper, the solid *M* used to obtain the Minkowski sum is a cube.

Figure 5 shows the nominal geometry of the tooth as it is used in the optimization process, together with offset geometries of different radii. Merging the data structures of the offset nominal geometry and the internal cavities allows for representing the complete (tessellated) boundaries of the simulation domain. Starting from this file, an appropriate tetrahedral mesh of the domain can be created with the aid of the CGAL library and its 3D meshing functionalities [27]. This tetrahedral mesh is then passed forward to the subsequent stage where the simulation of the printing process is performed.

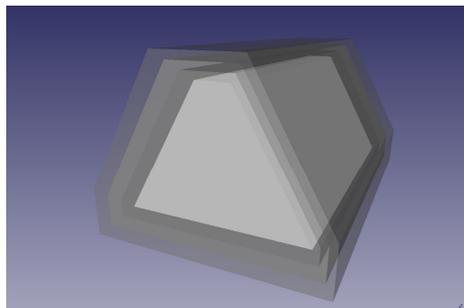

*Figure 5: Nominal geometry with different offset radii.*

### 2.5.3 Printing process simulation

The EOS M280 is the Selective Laser Sintering (SLS) printer considered in this work. The printing process consists of spreading uniformly the loose powder by a wiper to form a new layer ready for the laser sintering. A user defined scanning sequence drives the heat source (laser) to melt this new layer of the geometry of the component. The process is repeated until the stock part is built.

From the computational point of view, different approaches have been developed for the numerical simulation of this process. The most accurate one requires the solution of the fully coupled thermal and mechanical problem together with a high-fidelity representation of the scanning path [28] [11]. These approaches provide a reliable and accurate prediction of the distortion and residual stresses [29] [30] [31] but are expensive in terms of computational resources and CPU time. Nowadays, the numerical simulation of large industrial components is still unfeasible with standard computers and the use of massive parallel computing in distributed memory is mandatory [12] [32]. Consequently, the implementation of a simplified method was necessary to reduce the CPU time. In this work, the *Inherent Shrinkage method* has been adopted [33] [12] [32].

The main hypothesis allowing for the use of the Inherent Shrinkage Method is that it assumes a very localized heat source (laser). Hence, the melting process occurs exclusively at the Heat Affected Zone (HAZ) without affecting the remaining part of the domain. This given, the classical coupled non-linear thermo-mechanical analysis is skipped, and the original transient problem is replaced by a sequence of mechanical computations according to the deposition strategy.

More precisely, the distortion is assumed to be due to the shrinkage of the thermal-affected powder during the cooling/sintering phase; therefore, it is possible to predefine a strain tensor depending on the material parameters (melting temperature and thermal expansion coefficient), that takes into account both the thermal deformation and the plastic strains induced by the scanning sequence during powder sintering. The plastic strains have been calibrated and optimized according to the process parameters, namely the scanning speed and the power heat [33].

The computation is defined by adding new layers to the computational domain following the sequence of powder layers spread one after the other until completing the manufacturing of the full geometry. A further simplification consists of assuming a layer-by-layer deposition, skipping the high-fidelity simulation of the scanning sequence necessary to complete each layer of the domain. Hence, all the elements belonging to a new layer are *activated* to form part of the current computational domain. For each layer, the thermo-mechanical loading is given in terms of inherent strains at each Gaussian quadrature point element-by-element. Nonetheless, a huge number of layers (corresponding to an equal number of simulation steps) are necessary to deal with the numerical simulation. Indeed, in SLM technology each layer is about 20-30 microns thick, and the stock part can be a few millimeters (or centimeters) thick. As a result, an extremely fine mesh (smaller than the layer thickness) is also required increasing the CPU time per simulation. However, a sensitivity analysis to a multi-layer activation process has been also carried out, showing that it is possible to pack up to 10 layers per single activation without losing the original accuracy. Hence, up to 200 microns (or 0.2 mm) can be used as a reference thickness for the layer-by-layer analysis. The simulation time can be accelerated by a factor of 10 allowing the simulation kernel to be part of the optimization loop.

The other important hypotheses of the numerical simulation strategy are:

- the back plate is generally not analyzed. Instead, the corresponding clamping conditions are taken into account by the means of full displacement prescriptions at the contact surface between the part and the back-plate;
- the loose powder has been virtualized, thus it is neither discretized nor computed. This is due to the fact that, from the thermal point of view, the powder can be tackled as almost adiabatic boundary conditions, while mechanically, its stiffness is negligible if compared to the bulk (sintered material);
- the effects induced by the supporting structures have been accounted for. Their geometrical description as well their discretization by the FE mesh has been avoided to reduce the computational time while preserving the accuracy of the results. Instead, an equivalent stiffness has been computed and assembled to the global system of equations at each node connecting the part to the back-plate. This equivalent stiffness takes into account both the length and the effective area of each supporting structure as well as the fact that they are built by sintering the same powder as for the component. Hence, the Young's modulus of the supporting structure is the same as for the AM part under construction.

The final result coming from the manufacturing process simulation is presented in terms of accumulated distortions of the component according to the AM sequence. Those deformations will define a distorted volume to be compared to the nominal geometry. Comparisons with both the experimental evidence through the 3D printing of ad-hoc samples or by analyzing the results obtained by High-Fidelity fully coupled thermo-mechanical analysis have been carried out [33] [29] [31] [30].

The proposed AM model has been implemented into the software package COMET [34], a Finite Element based platform for the analysis of couple thermo-mechanical problems. Pre-and post-processing are done with the software GiD [35], developed at CIMNE - International Center for Numerical Methods in Engineering, www.cimne.com.

### 2.5.4 Computation of the distance between warped stock design and nominal geometry

To evaluate the constraint function of the optimization problem, it is necessary to compute the distance between the nominal geometry and each of the points belonging to the tessellation of the external surface of the distorted geometry. This computation is based on an iterative closest point projection method, see e.g. [36]. The method consists of an iterative solution to find the orthogonal projection of each point to the (smooth) surface defining the nominal geometry. The iterations are based on Taylor expansion of the expression to be minimized. If the closest point projection results in a point outside the domain then the distance is computed with respect to the closest boundary of the trimmed surface.

To improve the performance of the searching algorithm, enhanced topology information is used: for each surface belonging to the nominal geometry, all its neighboring surfaces are computed. Hence, if the closest point projection lays outside the original boundary surface, then the projection can be expected to be found in one of the neighboring surfaces. Moreover, the position of the points resulting from the previous projection is chosen as the best starting point for the next iteration. Figure 6 shows the position of the nodes belonging to the distorted mesh together with the nominal geometry.

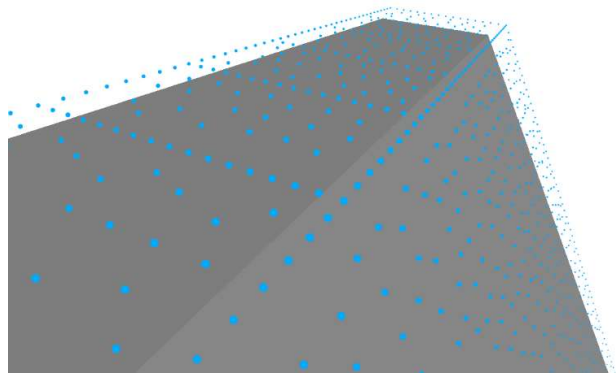

*Figure 6: Nominal geometry and points cloud used to measure the distance of each surface of the warped geometry.*

### 2.5.5 Computation of the objective function

The last step of the full model evaluation for a single stock part is the computation of the objective function $\Delta Volume$, which consists of the volume of the extra material to be removed from the stock part by machining. This volume can be obtained by subtracting the volume of the nominal geometry from the volume of the printed stock part. The volume of the nominal geometry can be computed once and for all beforehand; the volume of the printed stock part can be obtained by a post-process of the results of the AM process simulation, simply by summing the volumes of the tetrahedra of the warped mesh. Of course, the computational mesh of the AM process does not include the internal cavities, whose volume needs than to be added back:

$$\Delta Volume(\mathbf{p}) = \text{Warped mesh volume}(\mathbf{p}) + \text{cavities volume}(\mathbf{p}) - \text{nominal geometry volume}.$$

In principle, also the cavities volume needs to be computed after warping; however, we approximately consider the volume of the cavities to be not significantly affected by the warping (in other words, the warping is active only on the outer surface of the stock part). Therefore, the volume of the cavities can be computed before the AM process.

## 2.6 Implementation details

As described in the previous sections, the entire methodology consists of many steps that cover different aspects (optimization, geometry manipulation and queries, AM simulation, mesh generation, post-processing, etc.) and needs different pieces of software. One objective of the CAxMan project was the interoperability of the software involved in the AM industry through a cloud-based system. For this reason, this work was conceived as testing platform for a complex workflow in the CAxMan cloud system and to establish the proper interoperability between each piece of software, either by designing suitable API or by files exchange, favoring standard formats like STEP or devising format converters if a standard format was not suitable or usable by a specific software. Once the file information format was defined, we were able to run in a fully automatic way the entire methodology described above.

The software used was a mixture of third-party and in-house developed software, running in a Linux environment (using Docker containers to simplify the deployment of each component, https://www.docker.com). Running the individual pieces of software was performed by a bash script, which performs "breadth-wise" operations, i.e., it runs each piece of software over all stock-parts under investigation (possibly in parallel) and then moves to the next step. The script takes also care of communication between pieces of software (typically, passing as inputs to a software the files that were produced as output in the previous step).

Specifically:

- SG++ (http://sgpp.sparsegrids.org) was used for generating the sparse grids and solving the optimization problem (1.1)
- Internal cavities were produced with an in-house software;
- Applying offsets to the nominal geometry was performed by CGAL routines (https://www.cgal.org);
- Merging the representations of cavities and offsetted geometry, as well as some clean-up which was sometimes needed, was done by in-house software;
- Generating the mesh for the AM process was performed by CGAL;
- Simulating the AM process and computing the volume of the stock-part was performed by the in-house software COMET [34], while the post-processing was performed with the software GiD [35];
- Computing distances between warped stock-part and nominal geometry was performed by in-house software, adapted from the code previously developed within the CloudFlow project (https://eu-cloudflow.eu).

Figure 7 shows the flowchart of the implemented optimization methodology.

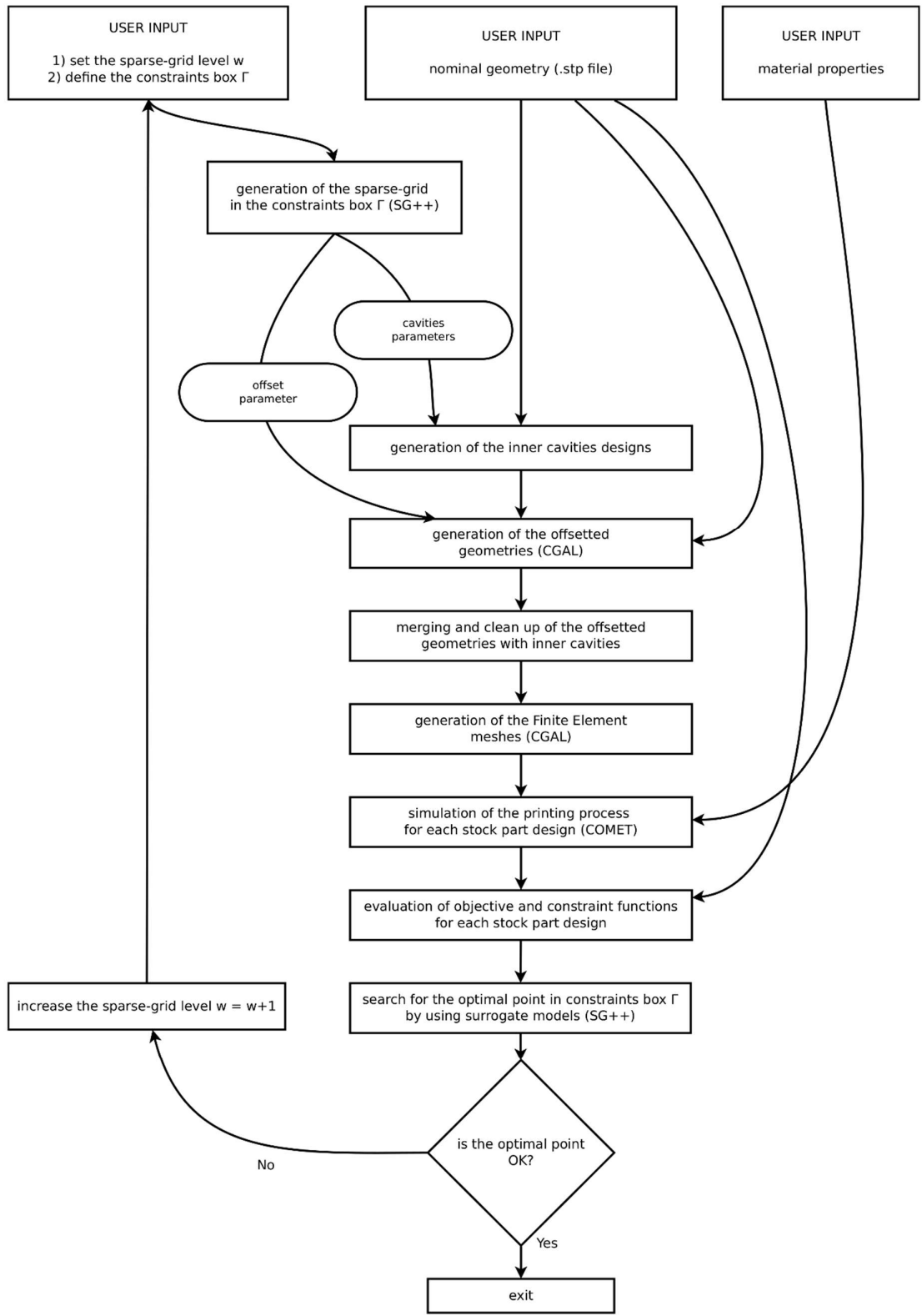

*Figure 7: Software flowchart for the optimization methodology implemented.*

## 3  Numerical Results

The results of the optimization methodology presented are discussed here. A sequence of tests of increasing complexity is considered, by extending the number of optimization parameters. Ti64 is the reference powder material used for the printing process. The following properties are required for the AM process simulation by inherent strain method:

- Density = 4420 kg/m$^3$
- Young Modulus = 1.18e+11 Pa
- Poisson Ratio = 0.33
- Thermal expansion coefficient = 9e-06
- Metal deposition temperature = 700 ºC
- Annealing temperature = 800ºC
- Elastic Limit = 954e+06 Pa
- Ultimate stress = 1110e+06 Pa

The nominal geometry is contained in a bounding box of 10 mm x 6 mm x 4 mm and the machining tolerance is set as 0.04 mm.

### 3.1  Optimization of the offset thickness

This analysis focuses on the optimization of the offset thickness only. The offset value ranges between 0 and 1 mm. The AM simulation uses a mesh defined by approximately 500K tetrahedra. Figure 8 shows the thermal shrinkage induced by the 3D printing process.

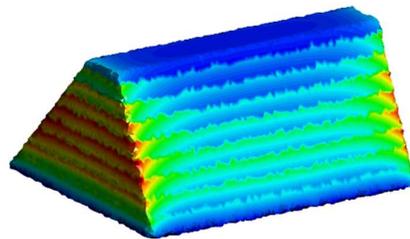

*Figure 8: Thermal shrinkage induced by the AM process. The scale ranges between 0 mm(blue) and 2mm (red).*

The optimization workflow is performed four times by increasing the sparse-grid level from *w*=2 to *w*=5, until reaching the required accuracy (i.e., the optimal thickness value computed for w=4 and w=5 differ less than 1e-4 mm). Figure 9 shows the surrogate models of the objective and constraint functions, respectively, for *w*=4: left figure for the objective and right figure for the constraint function. The constraint function is actually rewritten as $0.04mm - \Delta Thickness(\boldsymbol{p}) < 0$ due to requirements of the optimization software. The black triangles represent the values computed by the evaluation of the full model (i.e., performing all the steps described in Section 2.5), while the red dotted lines represent the surrogate functions computed out of those full model evaluations. As expected, the objective function (i.e., the volume) increases as the offset thickness increases, more than linearly. The distance between the nominal and the distorted surfaces also becomes larger and larger as the offset increase. Hence, the constraint function becomes more and more negative. Note that if the constraint function was positive, the offset would not be feasible because the machining tolerance would not be satisfied.

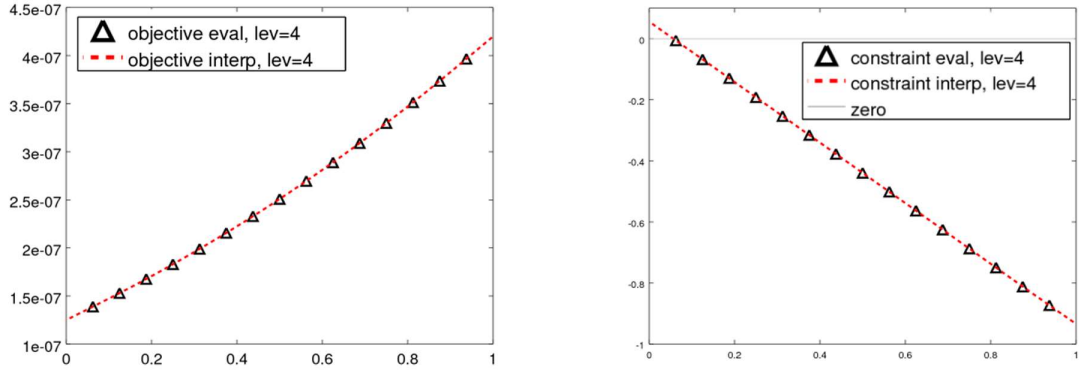

*Figure 9: surrogate model for objective (left) and constraint (right) functions for w=4 when optimizing the offset radius. The plot of the constraint function also reports the feasible region, which is the set of values of offset for which the constraint function is below the grey line.*

The results obtained through the optimization methodology are reported in Table 3. The column "interpolant evaluations" in Table 3 reports the number of evaluations of the surrogate models of the objective and the constraint functions performed by the optimization method to compute the optimal design point. These numbers would roughly correspond to the number of solutions required by the optimization process when sparse-grid surrogate model is not used. The value is significantly larger than the number of design points, meaning that a significant computational time is saved with the proposed procedure. This is even more true if one considers that this is the number of evaluations for a single run of the optimizer. In this work, the optimization procedure was repeated according to the 6 variants of the method and for 5 starting points to increase the likelihood of finding the exact optimal value of the offset.

| w | Design points | Optimal offset (mm) | Optimal volume (mm³) | Method | Interpolant evaluations | Computational time (hh:mm:ss) |
|---|---|---|---|---|---|---|
| 2 | 3 | 0.0569 | 132.67 | Squared penalty gradient | 52 | 00:12:01 |
| 3 | 7 | 0.0571 | 136.93 | Squared penalty gradient | 50 | 00:25:41 |
| 4 | 15 | 0.0567 | 137.59 | Squared penalty gradient | 49 | 00:56:35 |
| 5 | 31 | 0.0567 | 137.62 | Squared penalty gradient | 48 | 01:57:14 |

*Table 3: Results of the optimization test for offset radius only. Optimal volume measures the nominal geometry plus extra material. In the column method we report the method that delivered the best result, specifying both the penalization method (Squared Penalty, Augmented Lagrangian, Log-Barrier) and the unconstrained optimizer used (gradient, Nelder-Mead). Interpolant evaluations are only for the run of the best method which delivered the best result.*

## 3.2 Optimization of the offset thickness and one cavity design parameter

In this second test, two parameters are considered for the optimization: the offsetting value and one between the grid resolution and the minimum wall thickness used for the generation of cavity structures. The offset thickness ranges from 0mm to 0.1mm while the resolution for the cavity grid ranges from 17 to 24 and the wall thickness ranges from 0.4mm to 0.9mm. The number of tetrahedral elements used for the mesh discretization is the same as in the previous analysis. The sparse-grid level considered goes from *w*=2 to *w*=4 because *w*=5 results in an excessive number of points within parameter space and, thus, in the total computational time.

In Figure 10 and Table 4 the optimization of the offset thickness and the grid resolution is analyzed, with the second cavity design parameter kept fixed at its average value. In details, the left plot in Figure 10 shows with black dots the full model evaluation of the objective function, and the colored surface represents its sparse grids surrogate model. The red markers represent the best two solutions computed by the optimization procedure. The plot on the right shows instead the evaluation and sparse grids surrogate model of the constraint function, rewritten as in the previous case as $0.04 - \Delta thickness(\mathbf{p}) < 0$. The grey plane is set at height 0, therefore the feasible designs are those for which the surrogate model is below the grey plane. the offset thickness is the most important parameter: indeed, the surrogate models show a larger variation along the offset thickness axis than along the grid-resolution axis.

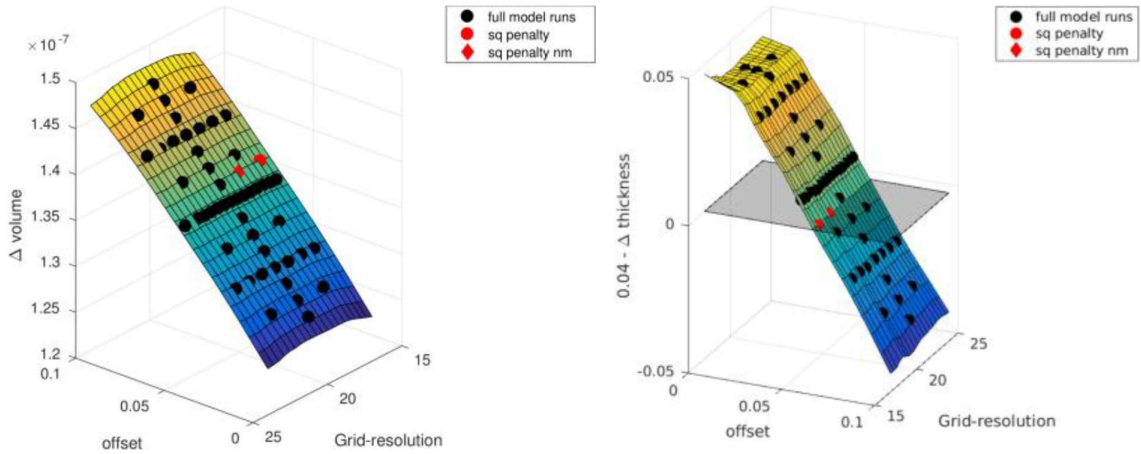

*Figure 10: surrogate models for objective and constraint functions for w=4 when optimizing the offset and the resolution of the void grid. A grey plane is used to indicate the region of admissibility (feasible region under the gray plane).*

Table 4 reports the results obtained by the optimization procedure. Because the offset thickness is the most important parameter, the optimal offset choice is similar to the one achieved in the previous section; the value of the optimal grid resolution is not constant as the sparse grid level increases but from the plots of the surrogate models we know that its value is not really relevant and could essentially be any admissible value. Nevertheless, it is still desirable to include cavities structures to enhance the thermal diffusion and to save printing time/material. Finally, observe the much larger number of surrogate model evaluations needed by the Nelder-Mead algorithm for this optimization case. This number is so large that trying to solve the optimization problem without replacing the full model with its surrogate model approximation would be impossible.

In Figure 11 and Table 5 the results of the optimization of the offset-thickness and the wall-thickness are shown. The figure and the table can be interpreted in the same way as above, and the results are equivalent.

| w | Design points | Optimal offset (mm) | Optimal volume (mm2) | Optimal grid resolution | Method | interpolant evaluations | Computational time (hh:mm:ss) |
|---|---|---|---|---|---|---|---|
| 2 | 5 | 0.0568 | 137.96 | 24 | Squared penalty Nelder Mead | 1715 | 00:32:43 |
| 3 | 17 | 0.0576 | 138.31 | 22.4 | Squared penalty Nelder Mead | 2184 | 01:50:18 |
| 4 | 49 | 0.0576 | 138.25 | 17.6 | Squared penalty gradient | 58 | 05:15:11 |

*Table 4: Optimizing offset and grid resolution. Optimal volume measures the nominal geometry plus extra material.*

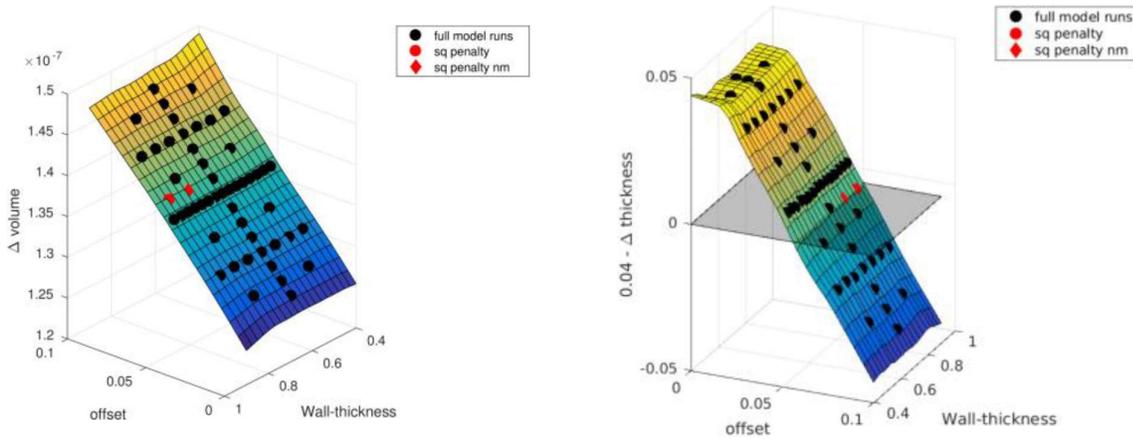

*Figure 11: surrogate models for objective and constraint functions for w=4 when optimizing the offset and the minimum wall thickness. A grey plane is used to indicate the region of admissibility.*

| w | Design points | Optimal offset (mm) | Optimal volume (mm³) | Optimal wall-thickness | Method | interpolant evaluations | Computational time (hh:mm:ss) |
|---|---|---|---|---|---|---|---|
| 2 | 5 | 0.0562 | 137.45 | 0.9 | Squared penalty Nelder Mead | 1731 | 00:22:34 |
| 3 | 17 | 0.0576 | 137.73 | 0.9 | Squared penalty Nelder Mead | 1733 | 01:18:51 |
| 4 | 49 | 0.0567 | 137.82 | 0.75 | Squared penalty Nelder Mead | 2200 | 03:47:49 |

*Table 5: optimizing offset and wall thickness. Optimal volume measures the nominal geometry plus extra material.*

### 3.3 Optimization of offset thickness and both cavity parameters

Finally, the most complex optimization analysis includes the offset thickness and the two parameters needed by the cavity generation algorithm. Here we limited ourselves to sparse grids of level w=2,3 because level w=4 would require a rather long computational time (approximately 6/7hours), without adding much value to the discussion. The variation ranges of the three parameters are the same as those used in the previous section. The results in this case are reported in Table 6 and are qualitatively similar to those found optimizing only offset thickness and one cavity design parameter: the optimal offset thickness is around 0.056mm, the optimal wall thickness is 0.9, i.e., as large as possible, while the optimal grid resolution is not constant across values of w. However, the fact that the optimal offset and the optimal volume is essentially unchanged from the previous tests suggests that once again, the parameters controlling the cavity generation do not affect significantly the optimal design of the offset material.

*Table 6: optimizing offset radius and both inner cavities generation parameters.*

| w | Design points | Optimal offset (mm) | Optimal volume (mm³) | Optimal grid resolution | Optimal wall-thickness | Method | Interpolant evaluations | Computational time (hh:mm:ss) |
|---|---|---|---|---|---|---|---|---|
| 2 | 7 | 0.0551 | 137.24 | 17 | 0.9 | Aug. Lagrangian Nelder-M. | 2008 | 00:34:35 |
| 3 | 31 | 0.0566 | 137.54 | 20.65 | 0.9 | Squared penalty Nelder-M. | 2057 | 02:27:23 |

# 4 Conclusions

The main objective of this work was the definition of the procedure for the optimization of the stock part design to be provided as input to a metal 3D-printer, keeping into account the thermal deformation induced by the AM process itself and the following post-production by subtractive machining. The problem has been defined as a multi-dimensional constrained optimization analysis, where the constraint and objective functions have been replaced by their sparse-grid surrogate models to achieve the best computational efficiency.

Even if the optimization of the stock part takes into account the generation of inner cavities, all the optimization analyses carried out indicate that the offset thickness has the largest impact on the distortion of the printed part. Hence, changing the parameters for the cavity creation does not significantly influence the heat-induced distortion.

The workflow presented can be further enhanced. In this work, the sparse grid level "*w*" is defined by the end-user according to the accuracy achieved. Alternatively, a further optimization loop could be added to keep increasing the sparse grid level until a satisfactory accuracy criterion is achieved; for instance, the optimization could stop when two consecutive executions of the optimization workflow result in the same optimal point (up to a certain tolerance).

From the modelling point of view, the offsetting is currently uniform on the whole surface of the nominal geometry. Using different offsets on different parts of the geometry could lead to a further improvement of the optimal solution. Moreover, the generation of inner cavities could be subjected to a mechanical constraint as a function of the performance of the component. In this case, not only the simulation of the manufacturing process but also the analysis of the mechanical/thermal performance of the component under the actual mechanical/thermal loading for each stock part design would be required as part of the optimization loop.

In addition to thermal distortions of the AM process, the post-processing phase also causes a change in the final shape because of the stress release due to the heat-treatment or the subtractive operations. Therefore, a numerical framework to simulate the entire manufacturing chain in AM could be used to predict the evolution of the part distortion and the corresponding residual stresses.

# 5 Acknowledgements

This work was funded by the European Union's Horizon 2020 research and innovation program under grant agreement No 680448 CAxMan, Computer Aided technologies for Additive Manufacturing.